\documentclass[runningheads]{llncs}

\usepackage[T1]{fontenc}

\usepackage[english]{babel}
\usepackage{subcaption}
\usepackage{url}
\usepackage{colortbl}
\usepackage{amssymb}
\usepackage{stmaryrd}
\usepackage{amsmath}
\usepackage{mathrsfs}
\usepackage{amssymb}
\usepackage[left]{lineno}
\usepackage{xfrac}
\usepackage{nicefrac}
\usepackage[textsize=scriptsize,backgroundcolor=yellow!40,obeyFinal]{todonotes}
\usepackage[nonumberlist,acronym,sanitize=none]{glossaries}
\glsdisablehyper
\usepackage{comment}
\usepackage[pdftex, colorlinks=true, hyperfootnotes=true, hyperindex=true,
             plainpages=false, pagebackref=false, pdfpagelabels=true, pdfstartview=FitH,
             linkcolor=purple, citecolor=teal, urlcolor=blue,
             bookmarks, bookmarksopen, bookmarksdepth=3]{hyperref}
\usepackage[capitalise,nameinlink]{cleveref}
\crefname{algocf}{alg.}{algs.}
\Crefname{algocf}{Algorithm}{Algorithms}
\crefname{section}{Sect.}{Sects.}
\Crefname{section}{Section}{Sections}
\usepackage{tkz-base}
\usetikzlibrary{decorations.pathmorphing,trees,snakes,arrows,shapes,automata,petri}
\usepackage{paralist}
\usepackage{multicol}
\usepackage{booktabs}
\usepackage{enumitem}
\usepackage{multirow}
\usepackage{rotating}
\usepackage{etaremune}
\usepackage[ruled,linesnumbered,algo2e]{algorithm2e}
\usepackage{marginnote}
\usepackage{mathtools}
\usepackage{adjustbox}
\usepackage{ifthen}
\usepackage[normalem]{ulem}
\usepackage{lineno}
\usepackage{soul}
\usepackage{floatrow}
\usepackage{listings}
\crefname{lstlisting}{Listing}{Listings}
\usepackage{lipsum}
\usepackage{makecell}
\usepackage{diagbox}
\usepackage[scientific-notation=false]{siunitx}
\usepackage{footmisc}
\usepackage{xspace}
\usepackage{ifdraft}
\usepackage{wrapfig}
\usepackage{orcidlink}
\newacronym[\glslongpluralkey={Distributed Ledger Technologies}]{dlt}{DLT}{Distributed Ledger Technology}

\newacronym{ipfs}{IPFS}{InterPlanetary File System}

\newacronym{p2p}{P2P}{peer-to-peer}

\newacronym{abe}{ABE}{Attribute-Based Encryption}
\newcommand{\ABE}[0] {\Gls{abe}\xspace}

\newacronym{maabe}{MA-ABE}{Multi-Authority Attribute-Based Encryption}
\newcommand{\MAABE}[0] {\Gls{maabe}\xspace}

\newacronym{cpabe}{CP-ABE}{Ciphertext-Policy Attribute-Based Encryption}
\newcommand{\CPABE}[0] {\Gls{cpabe}\xspace}

\newacronym{ssl}{SSL}{Secure Sockets Layer}

\newacronym{dht}{DHT}{Distributed Hash Table}

\newacronym{mpc}{MPC}{Multi-party Computation}

\newglossaryentry{box}{ %
  name={Box},
  description={authority intialisation storage box},
  first={\glsentrydesc{box} (henceforth, \glsentrytext{Box} for short)},
  plural={Boxes},
  firstplural={\glsentrydesc{box}es (henceforth, \glsentryplural{box} for short)}
}

\newglossaryentry{rloc}{ %
	name={resource locator},
	description={a generic term to denote content-based links obtained via hashing (e.g., the IPFS link)},
}

\newacronym{cpmaabe}{CP-MA-ABE}{Ciphertext-Policy Multi-Authority Attribute-Based Encryption}

\newglossaryentry{downer}{ %
  name={data owner},
  description={provide data for decision making},
  plural={data owners},
}

\def\DOwner{\glsentrytitlecase{downer}{name}\xspace}
\def\AttCert{Attribute Certifier\xspace}
\def\Reader{Reader\xspace}

\def\Auth{Authority\xspace}
\def\Auths{Authorities\xspace}

\newacronym{dk}{\textsl{dk}}{decryption key}
\newacronym{fdk}{\textsl{fdk}}{final decryption key}

\newacronym[\glslongpluralkey={Business Processes}]{bp}{BP}{Business Process}
\newacronym{bpi}{BPI}{Business Process Intelligence}
\newacronym{bpm}{BPM}{Business Process Management}
\newacronym{bpms}{BPMS}{Business Process Management System}
\newacronym{bpmn}{BPMN}{Business Process Model and Notation}
\newacronym{cpn}{CPN}{colored Petri net}
\newacronym{kpi}{KPI}{Key Performance Indicator}
\newacronym{ocbc}{OCBC}{Object-centric Behavioral Constraints}
\newacronym{soa}{SOA}{Service-Oriented Architecture}
\newacronym{pn}{PN}{Petri net}
\newacronym{wf}{WF}{workflow}
\newacronym{wfms}{WfMS}{Workflow Management System}
\newacronym{xes}{XES}{eXtensible Event Stream}
\newacronym{yawl}{YAWL}{Yet Another Workflow Language}
\newglossaryentry{task}{%
	name={task},description={the non-divisible, elementary activity}}

\newglossaryentry{promod}{%
	name={process model},description={the model of a process}
}
\def\LogAlph {\ensuremath{\Sigma}}
\newglossaryentry{logalph}{
	name={log alphabet},description={the process alphabet, as reflected in a log},%
	symbol={\LogAlph}}
\def\Evt {\ensuremath{e}}
\newglossaryentry{evt}{
	name={event},description={a record of an instantaneous fact during the process enactment},%
	symbol={\Evt}}
\def\Trc { \ensuremath{\tau} }

\newglossaryentry{trace}{
	name={trace},description={a sequence of \glsplural{evt}},%
	symbol={\Trc}}
\def\EvtLog {\ensuremath{L}}

\newglossaryentry{evtlog}{
	name={event log},description={a collection of \glstext{evttrace}s},%
	symbol={\EvtLog}}

\renewcommand\tabcolsep{5pt}
\newcolumntype{d}{>{\columncolor{gray!10}}c}
\newcolumntype{m}{>{\columncolor{gray!10}}l}
\newfloatcommand{capbtabbox}{table}[][\FBwidth]
\newenvironment{iiilist}%
{\begin{inparaenum}[\itshape(i)\upshape]}%
{\end{inparaenum}}

\RequirePackage{xparse}
\NewDocumentEnvironment{AuthNote}{+o+o}{%
	\IfValueT{#2}{\marginnote{\scriptsize{#2}}}%
	\begin{scriptsize}
		\colorbox{gray}%
		{\color{white} Note\IfValueT{#1}{ (#1)}:}%
		\quad%
		\color{brown}
}{%
	\normalcolor
	\end{scriptsize}
}

\RequirePackage{pifont}

\RequirePackage{lipsum}
\newcommand{\LipsumGray}[1][]{{\color{gray}\ifthenelse{\equal{#1}{}}{\lipsum}{\lipsum[#1]}}}

\RequirePackage{siunitx}
\newcolumntype{D}[1]{S[
	table-omit-exponent,
	round-mode=places,
	round-integer-to-decimal,
	round-precision={#1}]} 

\providecommand{\eg}{{e.g.,}\xspace}

\newcommand{\SmallCode}[1]{\footnotesize\texttt{#1}\normalsize}

\graphicspath{{./images/}}

\begin{document}

\title{MARTSIA: A Tool for Confidential Data Exchange via Public Blockchain}

\titlerunning{MARTSIA}

\author{
	Michele~Kryston\inst{1,2}\orcidlink{0009-0000-1491-2471}
	\and
    Edoardo~Marangone\inst{1}\orcidlink{0000-0002-0565-9168}
	\and
    Claudio~{Di~Ciccio}\inst{2}\orcidlink{0000-0001-5570-0475}
	\and\\
	Daniele~Friolo\inst{1}\orcidlink{0000-0003-0836-1735}
	\and
	{Eugenio~Nerio}~Nemmi\inst{1}\orcidlink{0000-0001-6518-7863}
	\and
	Mattia~Samory\inst{1}\orcidlink{0000-0002-4916-8352}
	\and\\
	Michele~Spina\inst{1}\orcidlink{0009-0003-6870-5525}
	\and
	Daniele~Venturi\inst{1}\orcidlink{0000-0003-2379-8564}
	\and
    Ingo~Weber\inst{3}\orcidlink{0000-0002-4833-5921}
}
\authorrunning{M. Kryston et al.}

\institute{
    Sapienza University of Rome, Rome, Italy\\
    \email{\href{mailto:kryston.1844733@studenti.uniroma1.it}{kryston.1844733@studenti.uniroma1.it}};
	\email{\href{mailto:edoardo.marangone@uniroma1.it}{edoardo.marangone@uniroma1.it}}
    %
	\and
    Utrecht University, Utrecht, the Netherlands,
	\email{\href{mailto:m.kryston@uu.nl}{m.kryston@uu.nl}}; 
    \email{\href{mailto:c.diciccio@uu.nl}{c.diciccio@uu.nl}}
    \and
	School of CIT, Technical University of Munich and Fraunhofer Gesellschaft, Munich, Germany,
    \email{\href{mailto:ingo.weber@tum.de}{ingo.weber@tum.de}}
}

\maketitle
\begin{abstract}
Blockchain technology streamlines multi-party collaborations in decentralized settings, especially when trust is limited or difficult to establish. While public blockchains enhance transparency and reliability by replicating data across all network nodes, they also conflict with confidentiality.
Here, we introduce Multi-Authority Approach to Transaction Systems for Interoperating Applications (MARTSIA) to address this challenge. MARTSIA provides fine-grained read-access control at the message-part level by combining user-defined policies with certifier-declared attributes. The approach guarantees that even though data is replicated across the network to maintain consistency, fault tolerance, and availability, its confidentiality is securely preserved through encryption.
To this end, MARTSIA integrates blockchain technologies, Multi-Authority Attribute-Based Encryption, and distributed hash-table file storages. This architecture effectively balances the transparency inherent in public blockchains with the privacy required for sensitive applications. We present the tool and its applicability in a business scenario.
\end{abstract}

\section{Introduction}
\label{sec:introduction}
The emergence of blockchain technology has reshaped secure and transparent interactions among untrusted parties~\cite{DBLP:conf/bpm/WeberXRGPM16}. This technology provides security, ensured by cryptography; resilience, achieved through the decentralization of network nodes; and transparency, as anyone can verify past transactions recorded in the ledger. Public blockchain protocols strengthen these guarantees at scale, as they resort to an open peer-to-peer network, where data is replicated across all nodes, improving its availability and integrity. 
In modern business environments, these blockchain properties offer significant advantages over traditional e-business models, which for decades have relied on centralized databases and intermediaries to safeguard trust and enforce processes.
Centralized systems excel in effectiveness although they suffer in several aspects such as single points of failure, susceptibility to data breaches and inefficiencies in data reconciliation between multiple parties.
Public blockchains typically entail non-negligible operational costs and additional processing times due to their decentralized nature, but these aspects are often compensated by the security and transparency they guarantee.
As a result, blockchain technology has experienced increasing adoption in information systems within enterprise domains over the last few years~\cite{StiehleW22}. 
Nevertheless, full data transparency may pose a problem in multi-party business settings wherein sensitive information needs to be kept confidential between involved participants
~\cite{DBLP:conf/bpm/KopkeN22,DBLP:journals/csur/ZhangXL19}.

To overcome this issue, we bridge the silos of blockchain technologies and encryption with information systems engineering.
In this paper, we present the tool implementation of Multi-Authority Approach to Transaction Systems for Interoperating Applications (MARTSIA)~\cite{DBLP:conf/edoc/MarangoneCFNVW23}, a 
fully decentralized framework providing confidential information sharing through public blockchain technologies, \MAABE, and distributed hash-table file storages.
Inspired by the Control Access via Key Encryption (CAKE)  approach~\cite{DBLP:conf/bpm/MarangoneCW22}, MARTSIA enhances decentralization by removing CAKE's central nodes needed for data encryption and decryption.

The paper proceeds as follows. In \cref{sec:background}, we discuss the underlying technologies of the framework, while \cref{sec:illustrativeExample} introduces a running example.
In \cref{sec:innovationsAndFeatures}, we present its features by depicting the application of MARTSIA in the running example. \Cref{sec:maturity} evaluates the maturity of the framework, assessing its versions and readiness for deployment. In \cref{sec:relatedWork}, we review related work in the literature, placing MARTSIA in the context of existing solutions. Finally, \cref{sec:conclusionAndFutureWork} concludes the paper and proposes directions for future research and development.

\section{Background}
\label{sec:background}
Before delving into our approach, in this section we discuss its building blocks.

A \textbf{blockchain} is a decentralized and distributed ledger that stores transactions securely and tamper-resistantly. Transactions are stored in blocks linked to the previous one, forming a chain. This structure ensures transparency, security, and immutability of recorded information.
Most blockchains, such as Ethereum,\footnote{\label{foot:ethereum} \url{https://ethereum.org/}, accessed: 2025-03-12.} are additionally featured with the ability to execute code, namely \textbf{smart contracts}, in a decentralized manner through virtual machines, thus allowing complex protocols to be built at an application layer. 

The cost of performing a transaction primarily depends on the amount of information being stored and, in the case of a smart contract, on the computational effort required for execution.
For this reason, users often leverage additional technologies, such as distributed hash-table file storages, to store large amounts of data. An example is \textbf{InterPlanetary File System} (\textbf{IPFS}).
As on public blockchains, data stored on IPFS are decentralized and transparent, making them accessible to anyone. IPFS allows data retrieval via content-addressed locators, \eg the hash of the stored data. In this way, a single-byte difference between two data generates a completely different locator. These locators can then be stored on the blockchain to benefit from its properties.

Access to data stored on IPFS can be restricted to specific users through the application of \textbf{\ABE}. \ABE, a public-key encryption scheme, links encrypted data with corresponding decryption keys via attributes. One type of \ABE is \CPABE, where each user is associated with attributes and data are encrypted using policies, \eg logical formulas built upon user attributes. However, \CPABE relies on a single authority to generate user decryption keys. \MAABE~\cite{DBLP:conf/tcc/Chase07} overcomes this limitation by improving decentralization. In \MAABE, multiple authorities generate partial decryption keys. The user who requests the partial keys can merge them, obtaining a complete key to decrypt the data.

\section{Running Example}
\label{sec:illustrativeExample}
\begin{figure}[t]
	\centering
	\includegraphics[width=0.9\textwidth]{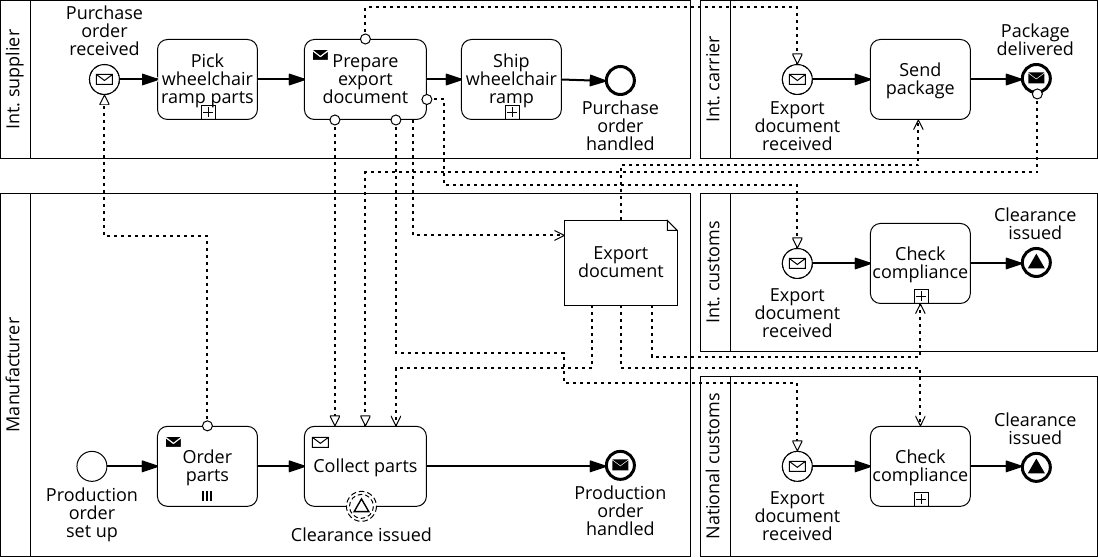}
	\caption{A segment of the collaboration diagram illustrating the control and data flow of the process}
	\label{fig:car-assembly-process}
\end{figure}
In this section, we introduce a running example that we will use throughout the paper to provide a clearer explanation of MARTSIA.

\Cref{fig:car-assembly-process} depicts a Business Process Model and Notation (BPMN) collaboration diagram illustrating a fragment of a supply-chain process in the automotive sector: the production of a custom car for an individual with paraplegia. 
In this fragment of the process, the Manufacturer proceeds to acquire the necessary ramp components for the cars if they are unavailable; in this situation, the Manufacturer orders them from an International supplier.
International customs supervises the international transit of goods, while National customs supervises the national transit. Both dispense Customs clearance. 
Thereafter, the International carrier ships the goods to the Manufacturer. 
We will focus on the Export document in \cref{fig:car-assembly-process}, which comprises multiple records, presented in \cref{tab:policies} as four distinct slices. Given a single sender, the International supplier, we have different recipients for each slice: the Manufacturer, the National customs, the International customs, and the International carrier. 

\section{Innovations and Features}
\label{sec:innovationsAndFeatures}
\begin{table}[t]
	\centering
	\resizebox{0.8\textwidth}{!}{%
		\arrayrulecolor[gray]{0.6} 

\begin{tabular}{|c|l|l|l|}
	\hline
	Slice & Recipients & Data & Policy \\ 
	\specialrule{0.08em}{0em}{0em} 
	1 & 
	\begin{tabular}[c]{@{}l@{}}Manufacturer\\ Nat.\ customs\\ Int.\ customs\\ Int.\ carrier\end{tabular} & 
	\begin{tabular}[c]{@{}l@{}}
\begin{lstlisting}[firstline=1,lastline=6]
Manufacturer:Beta
Delivery:    8, B Lane
E-mail:      b@mail.com
Ramp run:    3
Kickplate:   12
Amount_paid: $5000
\end{lstlisting} 
	\end{tabular} & 
	\begin{tabular}[c]{@{}l@{}}
\begin{lstlisting}[firstline=1,lastline=6]
Customs@A or
 ((Supplier@C and
   International@B) or
  Manufacturer@A or
  (Carrier@B and
   International@C))
\end{lstlisting}
	\end{tabular} \\ \hline
	2 & 
	\begin{tabular}[c]{@{}l@{}}Nat.\ customs\\ Int.\ customs\end{tabular} & 
	\begin{tabular}[c]{@{}l@{}}
\begin{lstlisting}[firstline=8,lastline=10]
Workers_rights:    Ok
Human_rights:      Ok
Ecosys_protection: Ok
\end{lstlisting} 
	\end{tabular} & 
	\begin{tabular}[c]{@{}l@{}}
\begin{lstlisting}[firstline=8,lastline=10]
Customs@A or
 (Supplier@D and
  International@B)
\end{lstlisting}
	\end{tabular} \\ \hline
	3 & 
	Manufacturer & 
	\begin{tabular}[c]{@{}l@{}}
\begin{lstlisting}[firstline=12,lastline=14]
Manufacturer:Beta
Address:     8, B Lane
Reference:   26487
\end{lstlisting} 
	\end{tabular} & 
	\begin{tabular}[c]{@{}l@{}}
\begin{lstlisting}[firstline=12,lastline=14]
(Supplier@2+ and
 International@B) or
  Manufacturer@A
\end{lstlisting}
	\end{tabular} \\ \hline
	4 & 
	\begin{tabular}[c]{@{}l@{}}Manufacturer\\ Nat.\ customs\\ Int.\ customs\end{tabular} & 
	\begin{tabular}[c]{@{}l@{}}
\begin{lstlisting}[firstline=16,lastline=20]
Invoice_ID:  101711
Billing:     5, G Lane
Gross_total: $5000
Company_VAT: U12345678
Issue_date:  2022-05-12
\end{lstlisting} 
	\end{tabular} & 
	\begin{tabular}[c]{@{}l@{}}
\begin{lstlisting}[firstline=16,lastline=19]
Customs@A or
 ((Supplier@C and
   International@B) or
  Manufacturer@A)
\end{lstlisting}
	\end{tabular} \\ \hline
\end{tabular}
	}
	\caption{The Export document in clear, sent by the International supplier}
	\label{tab:policies}
\end{table}
In this section, we explain the components and operations of MARTSIA considering the example described in \cref{sec:illustrativeExample}.

\begin{figure}[t]
	\centering
	\includegraphics[width=1\textwidth]{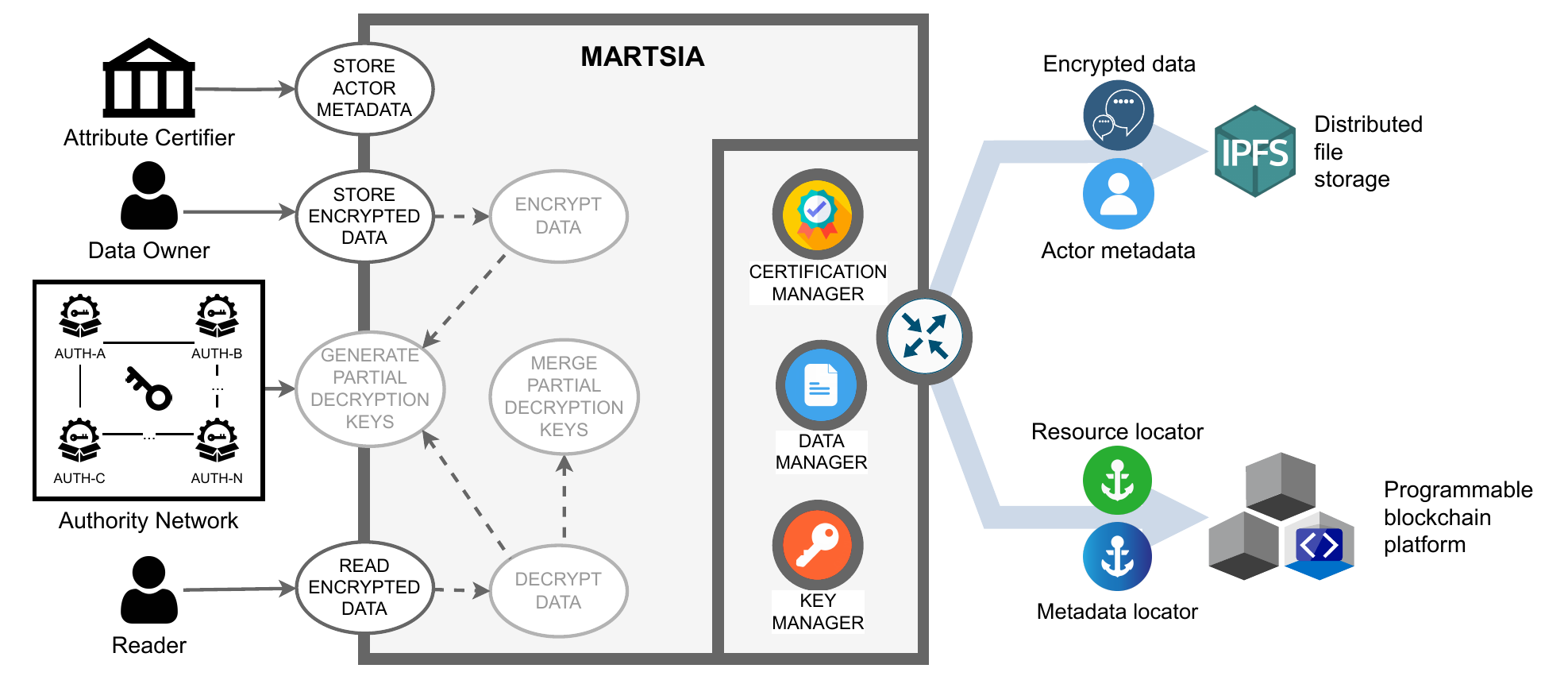}
	\caption[An overview of the MARTSIA architecture]{An overview of the MARTSIA architecture}
	\label{fig:martsia:architecture}
\end{figure}

\Cref{fig:martsia:architecture} outlines the general architecture of MARTSIA.
Our solution provides three main functionalities:
\begin{iiilist}
	\item \textbf{Store actor metadata}, assigning attributes to users. Considering our example, the chosen attributes designated to match the actors' names are: \texttt{Supplier}, \texttt{Manufacturer}, \texttt{Customs}, and \texttt{Carrier}. Furthermore, we introduce the \texttt{International} attribute. The International supplier, the International customs, and the International carrier are thus characterized by the conjunction of the \texttt{International} attribute with \texttt{Supplier}, \texttt{Customs}, and \texttt{Carrier}, respectively;
	\item \textbf{Store encrypted data}, applying a policy expressed as a propositional formula built upon users' attributes, thereby restricting access only to authorized actors. \Cref{tab:policies} outlines the policies governing read access to the Export document;
	\item \textbf{Read encrypted data}, accessing and decrypting data if the requester's attributes satisfy the access policy. In the example, column \textit{Data} in \cref{tab:policies} shows the decrypted data.
\end{iiilist}
These three functionalities include the support of other tasks (depicted in grey in the figure) we explain next.

Our solution employs three fundamental components:
\begin{iiilist}
	\item the \textbf{Certification Manager} records the attributes the users hold in a collaborative process;
	\item the \textbf{Data Manager} encrypts and stores data; and the
	\item \textbf{Key Manager} generates partial decryption keys derived from the user's attributes. The user merges these partial decryption keys into a single key to read encrypted data.
\end{iiilist}

Four main classes of actors are involved in the interaction:
\begin{iiilist}
	\item the \textbf{\AttCert} deploys the smart contracts, and assigns and stores the actors' metadata; 
	\item the \textbf{\DOwner} encrypts, determines reading access to the data, and stores it for sharing. In our example, this role is played by the International supplier, who encrypts the Export document;
	\item the \textbf{\Reader} interested in reading the shared data. In the example, this role is fulfilled by the International supplier and the Recipients in \cref{tab:policies};
	\item the \textbf{\Auth Network} generates partial decryption keys tailored to the Reader's attributes. Henceforth, we assume that the network consists of four {\Auths}: A, B, C, and D.
\end{iiilist}

\begin{sloppypar}
Due to the costs associated with storing data directly on the blockchain, MARTSIA stores on InterPlanetary File System (IPFS)\footnote{\url{https://ipfs.tech/}, accessed: 2025-03-12.} the encrypted data and the actors' metadata, along with additional information.
Smart contracts on the blockchain are employed to record IPFS locators, ensuring data integrity. Consequently, although anyone can access the public ledger, it cannot discern information about the encrypted data, the {\DOwner}s, or the {\Reader}s. 
Confidentiality with recipients is preserved by uniquely denoting each process instance with an identifier. This identifier, automatically added to the policies by MARTSIA, is used to characterize the actors involved, thereby limiting the range of potential {\Reader}s. The \DOwner must specify the \Auths responsible for validating each policy. Considering the running example in \cref{sec:illustrativeExample}, the complete access policy of the third slice in \cref{tab:policies} reads as follows: \SmallCode{(43175279@4+ and ((Supplier@2+ \ and \ International@B) \ or \ Manufacturer@A))}.
In this case: the process instance \texttt{43175279} needs to be verified by all the \Auths, the attribute \texttt{Supplier} by at least two, attribute \texttt{International} by \Auth B, and attribute \texttt{Manufacturer} by \Auth A. 
This policy specifies that the International supplier (the sender) and the Manufacturer involved in process instance 43175279 can access this slice.
A detailed explanation of the policies and their grammar can be found in~\cite{DBLP:journals/corr/abs-2308-03791}.
\end{sloppypar}

Following the encryption phase, the International supplier uploads the document to IPFS through MARTSIA. The corresponding IPFS resource locator (\eg \texttt{Qmb4Hz[...]CAd8iG}) is then stored on the blockchain and associated with a unique \emph{message ID}. Since the Export document includes multiple slices for various recipients, each record has also a unique \emph{slice ID}. 
To obtain the decryption key, a \Reader must request all the partial decryption keys to the \Auths. The \Auths generate these partial keys using the {\Reader}'s metadata retrieved from the blockchain. 
Then, the \Reader assembles the key parts, obtaining the final key necessary to read the encrypted data. To retrieve them, we have implemented secure channels resorting to TLS (direct) and RSA (via blockchain). 
In our example, if the Manufacturer intends to read the third slice of the Export document (\cref{tab:policies}), they must request the decryption keys to all the \Auths.
If the Manufacturer is involved in the process instance 43175279 and possesses the correct \texttt{Manufacturer} attribute, they meet the policy requirements for decryption. In contrast, the National customs, which only holds the \texttt{Customs} attribute, does not meet the access policy and is consequently unable to decrypt the content.
An actor lacking authorization from all the \Auths cannot decrypt the data. Although different \Auths collaborate in validating the policy's formulae, the record can be decrypted only by the requesting actor.

Here, we can derive a core feature of MARTSIA. Multiple {\Reader}s read distinct slices out of a single encrypted data object. Therefore, one message can be generated for multiple {\Reader}s interested in separate parts of it, rather than sending a copy for each such part, at the risk of potential inconsistency and communication overhead. In our example, we have one Export document sliced into four parts, accessible by four, two, one, and three participants, respectively. However, the message remains one, rather than being spread across ten replicas.

\section{Maturity}
\label{sec:maturity}
In this section, we describe the current state of the MARTSIA tool.
%

We implemented MARTSIA in two different variants, to show its platform-independence. Its smart contracts run on the Ethereum Virtual Machine (EVM) and the Algorand Virtual Machine (AVM). Smart contracts are written in Solidity v.~0.8.20 for EVM and PyTeal v.~0.20.0 for AVM. 
All other scripts are encoded in Python, including the integration with IPFS. 
The source code and documentation with implementation details are available at \href{https://github.com/apwbs/MARTSIA}{\nolinkurl{github.com/apwbs/MARTSIA}}. Among other things, the tool also includes 
additional features such as the possibility to use multiple Attribute Certifiers 
employing multi-signature operations,
and the customizability of the communication means to exchange partial decryption keys (either direct, via TLS channels, or indirect, via blockchain).
Furthermore, we include the input data, launching scripts, and experimental results of performance tests run on different EVM blockchains to verify transaction costs and latency, and a Wiki offering a step-by-step tutorial on system setup and operation. To demonstrate 
MARTSIA’s versatility across key domains, we integrated and tested our prototype with three publicly available DApps from different domains: 
NFT trading, supply-chain management, and retail~\cite{DBLP:journals/corr/abs-2308-03791}. 

The scripts and Wiki to reproduce this demo for the EVM version are available at
\href{https://github.com/apwbs/MARTSIA-Demo}{\nolinkurl{github.com/apwbs/MARTSIA-Demo}}.
For a brief video showcase, visit \href{https://www.youtube.com/watch?v=RAcifWw1_B0}{\nolinkurl{youtube.com/watch?v=RAcifWw1\_B0}}.

\section{Related Work}
\label{sec:relatedWork}

In recent years, blockchain technology has experienced  significant developments due to its distinctive properties, sparking growing interest in solutions to ensure data confidentiality.
Zou et al.~\cite{Zoul} propose SPChain, a system built upon blockchain to guarantee privacy and effortless sharing of medical data. This is achieved through a proxy re-encryption scheme and local databases of medical institutions.
Gan et al.~\cite{Gan} present a search method for encrypted medical data stored on a blockchain, resorting to an access control mechanism. Their architecture relies on: a consortium blockchain built on Ethereum for experimental analysis; several encryption methods such as probabilistic encryption and order-preserving encryption; and a non-relational database to store data. 
Remaining in the medical field, Miyachi and Mackey~\cite{Miyachi} introduce a framework for data privacy based on both on-chain and off-chain systems. They leverage consortium blockchains and asymmetric cryptography, while storing data off-chain. 
Unlike~\cite{Gan,Miyachi,Zoul}, MARTSIA operates on public blockchains to store IPFS locators, encrypts sensitive information with MA-ABE, and stores data on IPFS.

Yan et al.~\cite{YanGWZXH} illustrate a scheme providing fine access control with low computational consumption. Their work leverages blockchain to store metadata, IPFS to store data, and proxies for encryption and decryption. In addition, their scheme provides policy hiding and attribute revocation. However, two centralized entities are employed: the Proxy encryption server (ES) and the Proxy decryption server (DS). 
Wu et al.~\cite{WuZZGZZ} in their work present an efficient attribute-based encryption scheme to ensure privacy. Their method leverages the blockchain for non-repudiation and data integrity.
Feng et al.~\cite{FengWG} introduce a scheme utilizing blockchain for data storage; and identity-based encryption (IBE) and attribute-based encryption (ABE) for sharing and verifying the correctness of data. In their scheme, there is a centralized entity called Private Key Generator that takes care of the setup phase and attribute key generation.
In contrast to ~\cite{FengWG,WuZZGZZ,YanGWZXH}, our solution utilizes a decentralized attribute-based encryption scheme.

\section{Conclusion and Future Work}
\label{sec:conclusionAndFutureWork}
In this paper, we introduced MARTSIA, a framework leveraging blockchain and \MAABE to control data access in multi-party business scenarios. Our approach resorts to IPFS to store encrypted data and actors' metadata, and to smart contracts to link IPFS resource locators to the blockchain. MARTSIA guarantees access control, data reliability, immutability, and auditability in a decentralized manner. Areas for future development include data access revocation through InterPlanetary Name System (IPNS), automatic policy validation, conducting robustness field tests, and expanding the policy language to open the door to secure data calculations using TEEs.

\begin{credits}
	\noindent\parbox{\textwidth}{
	\subsubsection{\ackname}
	This work was partly funded by projects SERICS (PE00000014) under the NRRP MUR program funded by the EU-NGEU, PINPOINT (B87G22000450001) under the PRIN MUR program, and Health-e-Data, funded by the EU-NGEU under the Cyber~4.0 NRRP MIMIT programme.
}
	
\end{credits}

\bibliographystyle{splncs04}
\bibliography{bibliography}

\begin{thebibliography}{10}
\providecommand{\url}[1]{\texttt{#1}}
\providecommand{\urlprefix}{URL }
\providecommand{\doi}[1]{https://doi.org/#1}

\bibitem{DBLP:conf/tcc/Chase07}
Chase, M.: Multi-authority attribute based encryption. In: {TCC}. vol.~4392, pp. 515--534 (2007)

\bibitem{FengWG}
Feng, T., Wang, D., Gong, R.: A blockchain-based efficient and verifiable attribute-based proxy re-encryption cloud sharing scheme. Inf.  \textbf{14}(5), ~281 (2023)

\bibitem{Gan}
Gan, C., Yang, H., Zhu, Q., Zhang, Y., Saini, A.: An encrypted medical blockchain data search method with access control mechanism. Inf. Process. Manag.  \textbf{60}(6),  103499 (2023)

\bibitem{DBLP:conf/bpm/KopkeN22}
K{\"{o}}pke, J., Necemer, M.: Measuring the effects of confidants on privacy in smart contracts. In: {BPM} (Blockchain and {RPA} Forum). vol.~459, pp. 84--99 (2022)

\bibitem{DBLP:journals/corr/abs-2308-03791}
Marangone, E., {Di Ciccio}, C., Friolo, D., Nemmi, E.N., Venturi, D., Weber, I.: Enabling data confidentiality with public blockchains. CoRR  \textbf{abs/2308.03791} (2023)

\bibitem{DBLP:conf/edoc/MarangoneCFNVW23}
Marangone, E., {Di Ciccio}, C., Friolo, D., Nemmi, E.N., Venturi, D., Weber, I.: {MARTSIA:} enabling data confidentiality for blockchain-based process execution. In: {EDOC}. vol. 14367, pp. 58--76 (2023)

\bibitem{DBLP:conf/bpm/MarangoneCW22}
Marangone, E., {Di Ciccio}, C., Weber, I.: Fine-grained data access control for collaborative process execution on blockchain. In: {BPM} (Blockchain and {RPA} Forum). vol.~459, pp. 51--67 (2022)

\bibitem{Miyachi}
Miyachi, K., Mackey, T.: {hOCBS:} {A} privacy-preserving blockchain framework for healthcare data leveraging an on-chain and off-chain system design. Inf. Process. Manag.  \textbf{58}(3),  102535 (2021)

\bibitem{StiehleW22}
Stiehle, F., Weber, I.: Blockchain for business process enactment: {A} taxonomy and systematic literature review. In: {BPM} (Blockchain and {RPA} Forum). vol.~459, pp. 5--20 (2022)

\bibitem{DBLP:conf/bpm/WeberXRGPM16}
Weber, I., Xu, X., Riveret, R., Governatori, G., Ponomarev, A., Mendling, J.: Untrusted business process monitoring and execution using blockchain. In: {BPM}. vol.~9850, pp. 329--347 (2016)

\bibitem{WuZZGZZ}
Wu, A., Zhang, Y., Zheng, X., Guo, R., Zhao, Q., Zheng, D.: Efficient and privacy-preserving traceable attribute-based encryption in blockchain. Ann. des T{\'{e}}l{\'{e}}communications  \textbf{74}(7-8),  401--411 (2019)

\bibitem{YanGWZXH}
Yan, L., Ge, L., Wang, Z., Zhang, G., Xu, J., Hu, Z.: Access control scheme based on blockchain and attribute-based searchable encryption in cloud environment. J. Cloud Comput.  \textbf{12}(1), ~61 (2023)

\bibitem{DBLP:journals/csur/ZhangXL19}
Zhang, R., Xue, R., Liu, L.: Security and privacy on blockchain. {ACM} Comput. Surv.  \textbf{52}(3),  51:1--51:34 (2019)

\bibitem{Zoul}
Zou, R., Lv, X., Zhao, J.: {SPC}hain: Blockchain-based medical data sharing and privacy-preserving ehealth system. Inf. Process. Manag.  \textbf{58}(4),  102604 (2021)

\end{thebibliography}

\end{document}